\documentclass{sig-alternate}

\usepackage{times} 
\usepackage{amsmath}
\usepackage{amsfonts}
\usepackage{amssymb}
\usepackage{algorithm}
\usepackage{algorithmic}
\usepackage{graphicx}
\usepackage{graphics}
\usepackage{paralist}
\usepackage{xspace}
\usepackage[usenames,dvipsnames]{xcolor}
\usepackage{multirow}
\usepackage{pifont}
\usepackage[T1]{fontenc}
\usepackage{array}
\usepackage{url}
\usepackage{booktabs}
\usepackage{xfrac}
\usepackage{caption}
\usepackage{subcaption}
\usepackage{pbox}
\usepackage[auth-lg]{authblk}
\usepackage{epstopdf}

\newtheorem{observation}{Observation}

\newtheorem{problem}{Problem}

\newtheorem{definition}{Definition}

\newcommand{\hide}[1]{}

\newcommand{\bit}{\begin{compactitem}}
\newcommand{\eit}{\end{compactitem}}
\newcommand{\ben}{\begin{compactenum}}
\newcommand{\een}{\end{compactenum}}

\newcommand{\method}{\textsc{EdgeCentric}\xspace}

\newcommand{\ourtheory}{{\it formulation}}
\newcommand{\ouralgorithm}{{\it methodology}}
\newcommand{\ourpracticality}{{\it practicality}}

\definecolor{OliveGreen}{rgb}{0,0.6,0}

\newcommand{\cross}{\textcolor{red}{\ding{56}}}
\newcommand{\tick}{\textcolor{OliveGreen}{\ding{52}}}
\newcommand{\notsure}{\textcolor{BurntOrange}{\textbf{?}}}

\newcommand{\flipkart}{{\tt Flipkart}\xspace}
\newcommand{\amazon}{{\tt AmazonHPC}\xspace}
\newcommand{\swm}{{\tt SWM}\xspace}

\author[1]{Neil Shah}
\author[1]{Alex Beutel}
\author[1]{Bryan Hooi}
\author[2]{Leman Akoglu}
\author[3]{Stephan G\"{u}nnemann}
\author[4]{Disha Makhija}
\author[4]{Mohit Kumar}
\author[1]{Christos Faloutsos}
\affil[1]{Carnegie Mellon University, Pittsburgh, PA, USA\\ \texttt{\{neilshah,abeutel,bhooi,christos\}@cs.cmu.edu}}
\affil[2]{Stony Brook University, Stony Brook, NY, USA\\ \texttt{leman@cs.stonybrook.edu}}
\affil[3]{Technische Universit\"{a}t M\"{u}nchen, Munich, Germany\\ \texttt{guennemann@in.tum.de}}
\affil[4]{Flipkart, Bangalore, India\\ \texttt{\{k.mohit,disha.makhiji\}@flipkart.com}}

\begin{document}
%

\title{EdgeCentric: Anomaly Detection in Edge-Attributed Networks}

\maketitle

\begin{abstract}
Given a network with attributed edges, how can we identify anomalous behavior?  Networks with edge attributes are commonplace in the real world.  For example, edges in e-commerce networks often indicate how users rated products and services in terms of number of stars, and edges in online social and phonecall networks contain temporal information about when friendships were formed and when users communicated with each other -- in such cases, edge attributes capture information about how the adjacent nodes interact with other entities in the network.  In this paper, we aim to utilize exactly this information to discern suspicious from typical node behavior.  Our work has a number of notable contributions, including 
(a) \emph{\ourtheory}: while most other graph-based anomaly detection works use structural graph connectivity or node information, we focus on the new problem of leveraging edge information, 
(b) \emph{\ouralgorithm}: we introduce \method, an intuitive and scalable compression-based approach for detecting edge-attributed graph anomalies, and
(c) \emph{\ourpracticality}: we show that \method successfully spots numerous such anomalies in several large, edge-attributed real-world graphs,
including the Flipkart e-commerce graph 
with {\em over 3 million} product reviews between {\em 1.1 million} users and {\em 545 thousand} products, where it achieved 0.87 precision over the top 100 results.
\end{abstract}



\section{Introduction}

\begin{figure*}[t]
    \centering
    \begin{subfigure}[b]{0.33\textwidth}
        \centering
        \includegraphics[width=\textwidth]{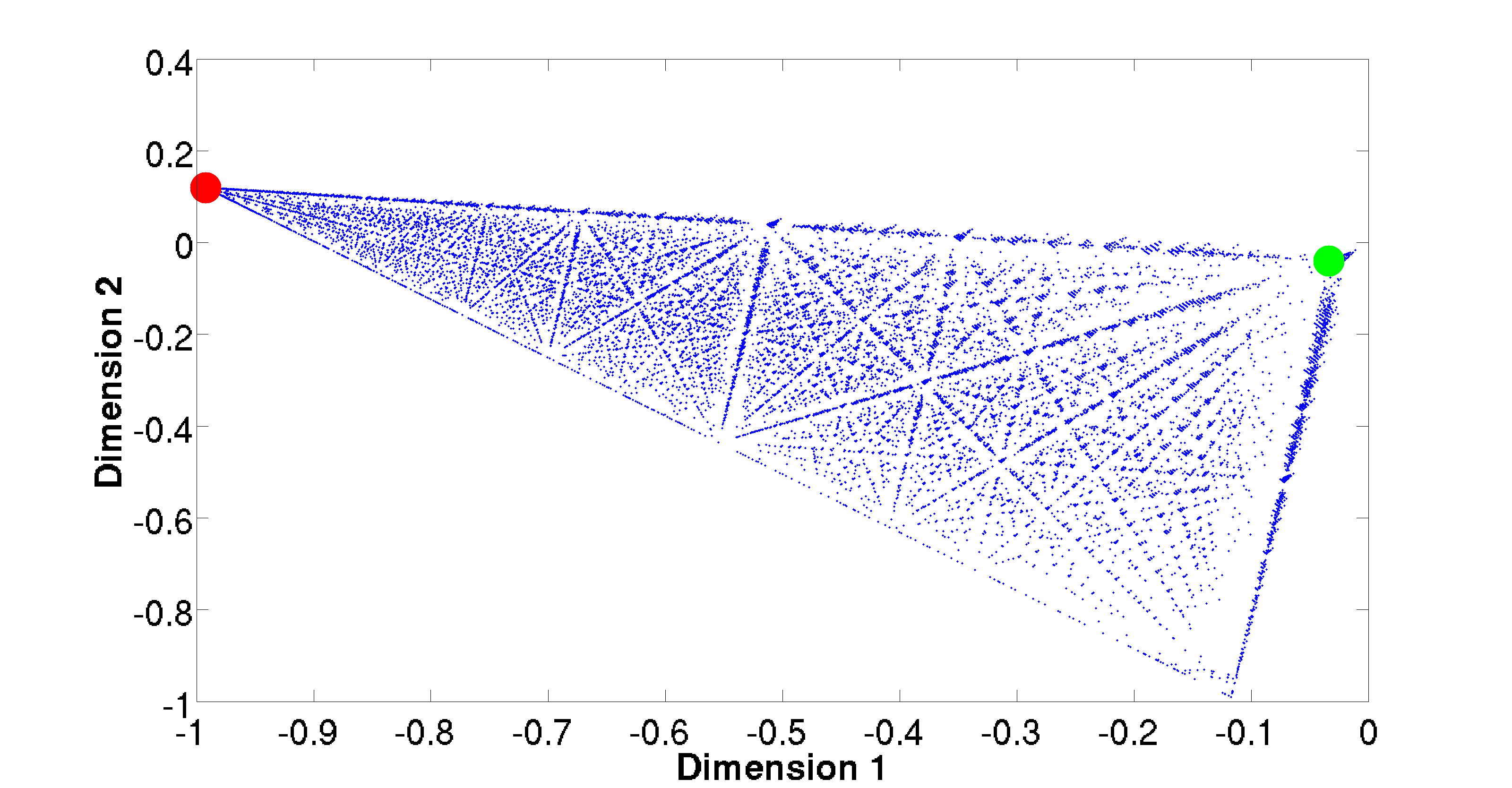}
        \caption{\label{fig:crown2d} Two clusters (red and green) of hard-to-discern fraudsters shown in a collapsed 2D subspace, reduced from the original 5D subspace over user rating values (1-5).}
    \end{subfigure}
    \hfill
    \begin{subfigure}[b]{0.33\textwidth}
        \centering
        \includegraphics[width=\textwidth]{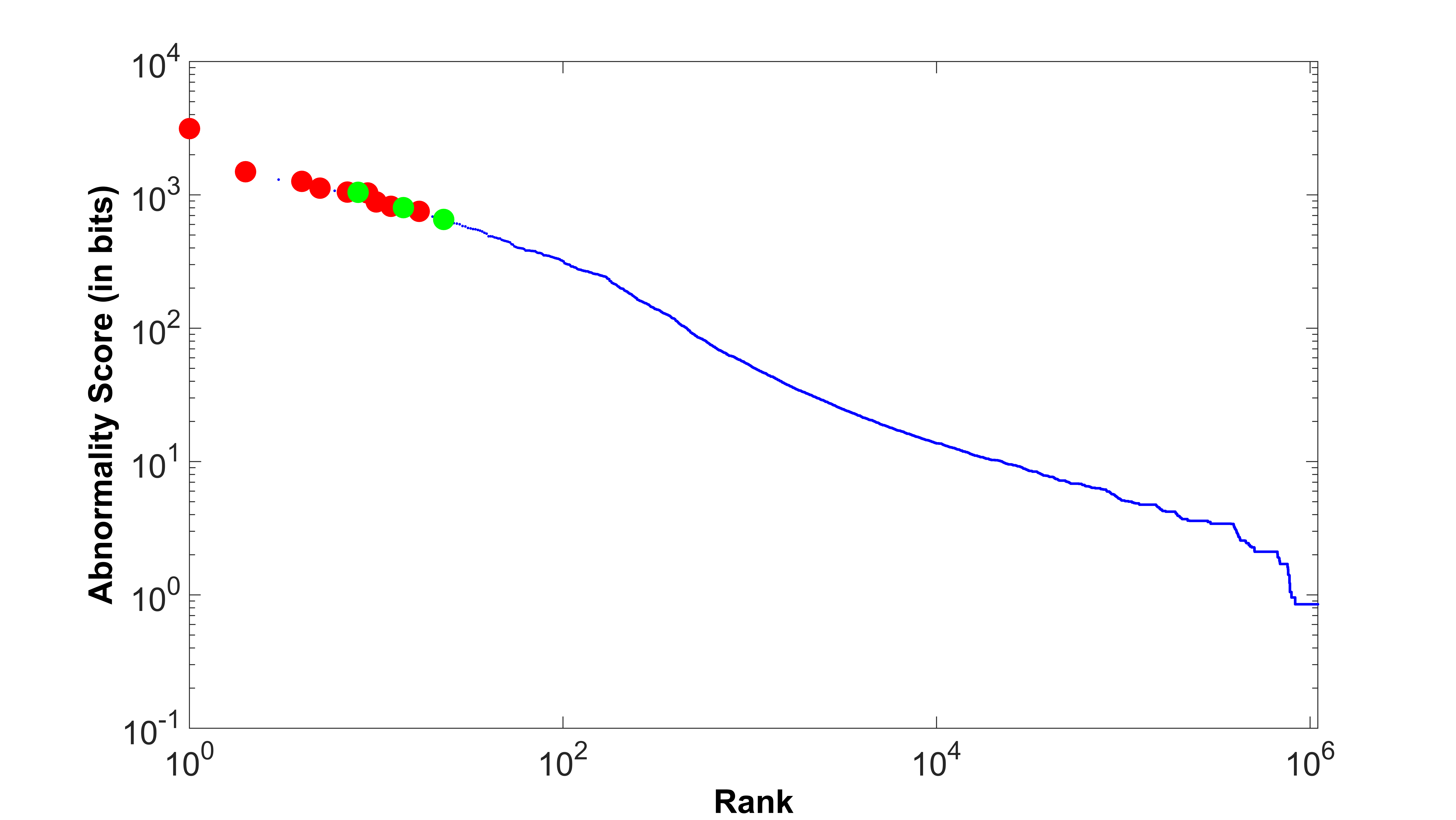}
        \caption{\label{fig:crownklv} Our approach, \method, identifies the users at the red and green clusters as highly abnormal.}
    \end{subfigure}
    \hfill
    \begin{subfigure}[b]{0.33\textwidth}
        \centering
        \includegraphics[width=\textwidth]{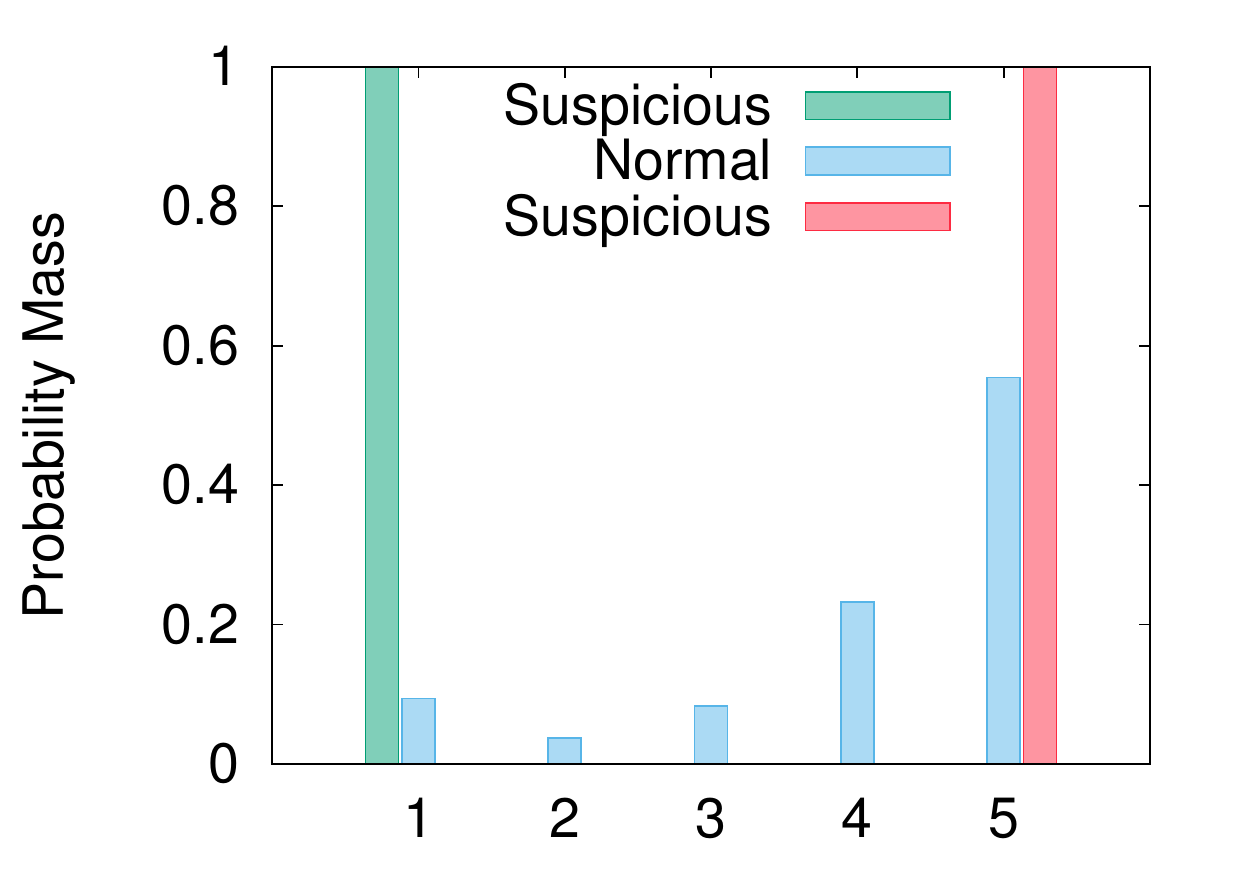}
        \caption{\label{fig:crownratings} We find that the abnormal users in the red cluster give only 5 star ratings, whereas users in the green cluster give only 1 star ratings.
	}
    \end{subfigure}
    \caption{\label{fig:crown} {\bf \method spots abnormal users on real graphs.}  
    Applied on a dataset of 3 million Flipkart user-product ratings, \method 
    finds users who greatly deviate from typical behavior -- the red and green clusters contain single-mindedly ``enthusiastic'' and ``disgusted'' users who only give 5 star or 1 star reviews respectively, compared to the global ($J$-shape) behavior shown in blue.}
\end{figure*}

Given a graph with attributed edges, what can we say about the behavior of the nodes?  For example, in a user-product graph with a rating attribute (1-5 stars) on edges, how can we discern which users are normal or abnormal? Furthermore, between two users with varying edge behavior, can we say which is more suspicious?  These are exactly the questions we address in this paper -- more specifically, we focus on the problem of leveraging edge-attributes in social and information graphs for anomaly detection and user behavior modeling purposes.

Informally, our problem is as follows:
\begin{problem}[Informal] 
{\bf Given} a static graph with possibly multiple numerical or categorical edge attributes, {\bf identify} the nodes with most irregular (adjacent) edge behavior in a {\bf scalable} fashion.
\end{problem}

This problem has numerous applications -- graphs with edge attributes are ubiquitous in the real-world.  Typically, these attributes take the form of numerical or categorical features which describe details about the interactions between two connected nodes. For example, edges in unipartite social graphs (e.g. Facebook, Twitter) may be attributed with temporal information indicating the beginning of a friend or follower relationship.  Similarly, in a who-calls-whom phone-call network (e.g. Sprint, Verizon), each caller-callee edge can be attributed with a timestamp and duration indicating when the call was made and how long it lasted.  Edge attributes allow for richer representations of interactions in heterogeneous (multi-partite, multi-relational) graphs as well -- for example, number-of-star ratings in user-product graphs (e.g. Amazon, Yelp) or play counts in user-media networks (e.g. Youtube, Spotify).  

In this work, we propose \method, an effective information theoretic approach for general node-based anomaly detection in edge-attributed graphs.   Specifically, our method leverages MDL (Minimum Description Length) to spot nodes with atypical edge-attribute behavior on adjacent edges in an unsupervised fashion.  Figure \ref{fig:crown} shows one application of \method on the Flipkart e-commerce network, where it is able to spot fraudulent users giving too many atypical rating values.  Figure \ref{fig:crown2d} shows a collapsed 2-dimensional subspace of users produced from the original 5-dimensional rating space (users rate products from 1-5 stars) which spectral algorithms or practitioners may examine in an effort to identify anomalous behavior.  In this space, we do not find any apparent, suspicious microclusters of abnormal users.  However, Figure \ref{fig:crownklv} shows that our \method approach successfully identifies the highly abnormal behaviors of users who give many ratings of only 5 stars (red 
cluster) or 1 stars (green cluster).  These behaviors deviate substantially from the normal (global) user behavior, shown as the blue $J$-shape in Figure \ref{fig:crownratings}.  

The main contributions of our work are as follows:

\begin{compactenum}
 \item {\bf Formulation:} We formalize the problem of anomaly detection on edge-attributed graphs using an information-theoretic approach. 
 \item {\bf Methodology:} We develop \method, an effective and scalable algorithm for the same.
 \item {\bf Practicality:}  We experiment with our \method on multiple 
 large, real-world graphs 
 and demonstrate its effectiveness and generality.
\end{compactenum}

\textbf{Reproducibility}: Our code for \method is open-sourced at \url{www.cs.cmu.edu/~neilshah/code/edgecentric.tar}.

\section{Related Work}

\begin{table*}[t!]
\centering
\caption{Feature-based comparison of \method with alternative approaches ($\notsure$ indicates limited support).}
\label{tbl:comp}
    \begin{tabular}{ccccc}
    \toprule
     & {\bf Prior-free} & {\bf Heterogeneous} & {\bf Independent edge-attributes} & {\bf Identifies anomalies} \\ \midrule
     \textsc{nfs} \cite{ye2015discovering} & \tick & \tick & \cross & \tick \\ 
     \textsc{fbox} \cite{shah2014spotting} & \tick & \tick &  \cross & \tick \\ 
     \textsc{catchsync} \cite{jiang2014catchsync} & \tick & \cross & \cross & \tick \\ 
     \textsc{copycatch} \cite{beutel2013copycatch} & \tick & \tick & \notsure & \tick \\
     \textsc{fraudeagle} \cite{akoglu2013opinion} & \cross & \tick & \notsure & \tick \\ 
     \textsc{netprobe} \cite{pandit2007netprobe}  & \cross & \tick & \cross & \tick \\ 
     \textsc{oddball} \cite{akoglu2010oddball} & \tick & \tick & \cross & \tick \\ 
     \textsc{crossspot} \cite{crossspot} & \tick & \tick & \notsure & \tick \\ 
	 \textsc{birdnest} \cite{birdnest} &  \cross & \tick & \notsure & \tick \\  
     \textsc{mimag} \cite{boden2012mining}, \textsc{rmics} \cite{boden2013rmics} & \tick & \cross & \tick & \cross \\ 
     \textsc{metis} \cite{karypis1995metis}, \textsc{graclus} \cite{dhillon2005fast}, \textsc{eigenspokes} \cite{prakash2010eigenspokes} & \tick & \cross & \cross & \cross \\  \midrule
     \method & \tick & \tick & \tick & \tick \\ \bottomrule
    \end{tabular}
\end{table*}

Prior work loosely falls into three categories:
(a) mining unattributed, or plain graphs, in which analysis is conducted using only connectivity information, 
(b) mining graphs with node attributes, which involves graphs for which descriptive features are placed on nodes, and 
(c) mining graphs with edge attributes, which focuses on graphs with features placed on edges.  
We describe the relevant work from each category next.
Table \ref{tbl:comp} gives a comparative analysis of existing methods, showing that none of the existing works satisfies all the relevant criteria for our problem setting.

\subsection{Mining unattributed graphs}
Akoglu et al.~\cite{akoglu2010oddball} 
identify power-law patterns in egonets and report deviating nodes as anomalous. 
Tong et al.~\cite{tong2011non} present a non-negative residual 
matrix factorization 
method to improve graph anomaly detection in low-rank subspaces.  
\cite{shah2014spotting, jiang2014inferring, jiang2014catchsync} 
propose spectral methods
to spot fraudulent behavior in low-rank subspaces 
of the Twitter and Weibo social network graphs.  
\cite{ghosh2012understanding} proposes a modified PageRank measure which penalizes fraudsters based on social linking promiscuity and collusion.
\cite{pandit2007netprobe} and
\cite{akoglu2013opinion} 
use belief propagation to spot fraudsters, 
on eBay, and on product-review sites, respectively.
\cite{ye2015discovering} proposes the \emph{network footprint score} 
to spot opinion spammers, exploiting self-similarity and neighborhood diversity. 

Methods for community detection, graph partitioning, and subgraph mining are also relevant, given that fraudulent and abnormal behavior in social networks often forms dense blocks. Numerous methods exist for graph partitioning, including the seminal \textsc{metis} algorithm~\cite{karypis1995metis}, as well as spectral methods \cite{white2005spectral, prakash2010eigenspokes}.  While many of these approaches non-trivially suffer from the difficulty of choosing the number of communities, several information-theoretic approaches have been proposed in response, including automatic cross-associations \cite{chakrabarti2004fully}, \textsc{VoG}~\cite{koutra2014vog} for static graphs and \textsc{TimeCrunch}~\cite{shah2015timecrunch} for time-evolving graphs.

\subsection{Mining graphs with node attributes}
\cite{gao2010community} unifies structural and attribute similarity 
and infers community and outliers using hidden Markov random fields. 
\cite{perozzi2014focused}  introduces a local, ``focused'' clustering approach which identifies clusters and cluster outliers given a set of seed nodes from which user interests are learned.  \cite{noble2003graph} proposes an MDL formulation for identifying both anomalous and common/recurrent graph substructures.  

\cite{hanisch2002co} proposes a variant of hierarchical average linkage clustering for coupled analysis of attributed gene expression and biological networks.  \cite{long2006spectral} uses spectral clustering to group various types of homogeneous node-attributed relational data.  \cite{gunnemann2010subspace} introduces a pruning-based algorithm to identify subspace clusters which also exhibit strong graph connectivity. \cite{akoglu2012pics} proposes an MDL formualtion for jointly reordering connectivity and feature matrices to identify attributed clusters.  \cite{zhou2009graph} proposes the use of a learnable, unified distance measure to weight the contributions of graph structure and node attribute similarity in clustering tasks.

\subsection{Mining graphs with edge attributes}

Very little work has been done on the topic of mining general edge-attributed graphs.  
In some cases, edge labels are construed as weights, which can be used by some cut-based \cite{aggarwal2010managing} and spectral clustering \cite{long2006spectral} approaches.  In our problem setting, we rather consider each edge as an interaction and each label (possibly numerous labels on each edge) as a \emph{feature} describing the interaction, which can be categorical or numerical and need not conform to the notion of edge weight.  

The recommendation systems community has also focused on learning models of graphs
with ratings \cite{koren2009matrix}, and in some cases these models have been
used to find outliers \cite{beutel2014cobafi}.
A related line of research involves mining online reviews, which can be
considered as textual edge attributes \cite{hu2004mining}.  Further work has focused
on linguistic indicators of fraud in online reviews \cite{jindal2008opinion}.

\cite{de2010surprising} introduces a truncated log-logistic model for call duration in phone-call networks.  \cite{beutel2013copycatch} uses local graph search on the Facebook user-likes-page graph with temporal edge features to find dense subgraphs within fixed timespans indicative of fraud.  \cite{boden2012mining, boden2013rmics} propose methods for mining dense subgraphs with similar subsets of attributes.  \cite{crossspot} formulates a related metric of suspiciousness but is based on Poisson distributions and thus limited to simple count data.  \cite{birdnest} also approaches the problem of modeling the distribution of ratings and interarrival times from a Bayesian perspective.  Our work differs in that it takes a frequentist approach based on MDL and is designed to handle any set of edge-attributes on complex heterogeneous graphs.

\vspace{1mm}
\noindent Overall, none of the existing works matches all the desirable features in Table~\ref{tbl:comp}.  Our proposed \method approach (a) needs no priors or existing node labels, (b) extends naturally to heterogeneous networks with multiple object and relation types, (c) supports multiple, independent edge-attributes and (d) can identify and rank anomalies.

\section{Problem Formulation}

In this section, we outline the first core contribution of our work: specifically, we formalize the problem of detecting anomalous nodes in networks using edge attributes by leveraging a compression paradigm, based on MDL.  For clarity, see Table \ref{tbl:symb} for an overview of the recurrent symbols used in the future discourse.  

\subsection{Preliminaries}

The Minimum Description Length (MDL) principle states that given a family of models $\mathcal{M}$, the best model $M\in\mathcal{M}$ for some observed data $\mathcal{D}$ is the one which minimizes the sum $L(M) + L(\mathcal{D}|M)$, where $L(M)$ is the description length in bits used to describe the model $M$, and $L(\mathcal{D}|M)$ is the description length in bits used to describe the data $\mathcal{D}$ encoded using the given model $M$.  MDL enforces lossless encoding to fairly evaluate various models.  In this paper, rather than using MDL to \emph{find the best model} for our given data, we instead use it to answer the question of \emph{how well the data fits} a given model.  The intuition behind this approach is that while data which fits the model well enjoys high compression and shorter resulting description length, data which is ill-represented by the model is compressed poorly and costs many more bits to encode.  

In our problem setting, we are given a static directed or undirected multigraph $G(\mathcal{V},\mathcal{E},m)$ in which nodes are connected by (possibly multiple) edges. Technically, $m:\mathcal{E}\rightarrow \{ \{u,v\} \mid u,v\in \mathcal{V} \} $ assigns each edge $e\in \mathcal{E}$ to a pair of nodes. Furthermore, we have object type and relation/edge type mapping functions $\Phi: \mathcal{V} \rightarrow \mathcal{B}$ and $\Psi: \mathcal{E} \rightarrow \mathcal{R}$, where each node $v \in \mathcal{V}$ is characterized by an object type $\Phi(v) \in \mathcal{B}$ and edge $e \in \mathcal{E}$ is characterized by a relation type $\Psi(e) \in \mathcal{R}$.  Here, we define an object type to reflect a node ``role,'' -- for example, a user or product.  A relation type reflects the relationship between two objects -- for example, user-rates-product.  
 When $|\mathcal{B}| = 1$ and $|\mathcal{R}| = 1$, the graph is \emph{homogeneous}; otherwise, it is \emph{heterogeneous}. Furthermore, edges of each relation $r \in \mathcal{R}$ are labeled with values corresponding to the same finite subset of numerical or categorical attributes chosen from attribute set $\mathcal{A}$, given by the mapping $\Omega: \mathcal{R} \rightarrow 2^{\mathcal{A}}$, where $2^{\mathcal{A}}$ denotes the power set of $\mathcal{A}$.  In other words, the graphs we consider can have numerous relation types, and edges of 
each relation type are characterized by a fixed number of the same attributes (at least 1).  In the remainder of the problem formulation, let us consider a simple, undirected user-product 
graph, in which $|\mathcal{B}| = 2$ (user objects, product objects) and $|\mathcal{R}| = 1$ (user-rates-product, or product-rated-by-user relation) for ease of explanation.  Let us also assume that we have only one attribute on the edges: say, rating of the product in terms of number of stars (1-5).   

Then, our formal problem definition is as follows:

\begin{table}[t!]
\scriptsize
\caption{Frequently used symbols and definitions}
\label{tbl:symb}
    \begin{tabular}{ll}
    \toprule
    {\bf Symbol}               & {\bf Definition}                                                                                                                         \\ \midrule
    $G$              & static input graph                                                                       \\
    $\mathcal{V}, |\mathcal{V}|$                    & node-set, \# of nodes of $G$ resp.                          \\ 
    $\mathcal{E}, |\mathcal{E}|$			& edge-set, \# of edges of $G$ resp. \\ 
    $m(\cdot)$ & function to realize the multi-graph \\
    $\mathcal{A}, |\mathcal{A}|$			& attribute-set, \# of total attributes across edges in $\mathcal{E}$ resp. \\
    $\mathcal{B}, \mathcal{R}$				& set of object types and relation types resp. \\ 
    $\Psi(\cdot)$	&	maps nodes in $\mathcal{V}$ to object types in $\mathcal{B}$ \\
    $\Phi(\cdot)$	&	maps edges in $\mathcal{E}$ to relation types in $\mathcal{R}$ \\
    $\Omega(\cdot)$	&	maps relation types in $\mathcal{R}$ to attribute sets in $2^{\mathcal{A}}$ \\
    $\delta(\cdot)$				& unified abnormality function, defined on nodes in $\mathcal{V}$\\ 
    $U, P$		& user, product resp. \\
    $C, C(i)$		& global (model) dist. $C$, prob. mass of $i$th element resp. \\
    $C_{u,r,w,j}, C_{p,r,w,k}$	& $j$th ($k$th) $U$ ($P$) model dist. on attr. $w$ and rel. $r$ resp. \\
    $\rho_{u,r,w,j}, \rho_{p,r,w,k}$ & $j$th ($k$th) $U$ ($P$) cluster prop. on attr. $w$ and rel. $r$ resp. \\
    $\hat{U}, \hat{P}$	& discrete prob. dist. (of ratings) for $U$ and $P$ resp. \\
    $f_{u,r}, f_{p,r}$	& rating vectors for $U$ and $P$ on relation $r$ resp. \\
    $H(\cdot)$	& Shannon entropy in bits, defined on discrete prob. dist. \\
    $KL(\cdot \parallel \cdot)$	& KL divergence in bits, defined on two discrete prob. dists. \\
    $M$		& data model $M$ \\
    $L(U,M)$				& \# of bits used to encode $M$ and $U$'s behavior given $M$ \\ 
    $L(M)$				& \# of bits to encode $M$ \\ \bottomrule
    \end{tabular}
\end{table}

\begin{problem}[Formal] 
{\bf Given} a static multigraph $G(\mathcal{V},\mathcal{E},m)$ with $\geq 1$ numerical or categorical edge attributes chosen from $\mathcal{A}$, {\bf devise} an abnormality function $\delta( \cdot )$ to score each node $v \in \mathcal{V}$ based on its (adjacent) edge attribute behavior, and {\bf identify} the most irregular nodes in a {\bf scalable} fashion.
\end{problem}

\subsection{Intuition}

In order to see how we can leverage an information theoretic perspective and use MDL to inspire the formulation of $\delta$, we must first consider our model and data representations.  In this regard, to encode each user node, we must store information about the user's associated interactions through edges.  In our running example, because each edge simply contains information about a single categorical attribute value (1-5), we must encode the attribute value to losslessly reconstruct the vector which describes the user's rating behavior.  Thus, for each user node, we will encode a vector of rating values, e.g. $[5, 5, 1, 2, 5, 3, \ldots]$.  
Likewise, to encode a product node, we store information about the product's associated interactions through edges.  Thus, we store the ratings that the product was given by various users: say, $[1, 2, 1, 3, 2,$ $2, \ldots]$. 

To encode these individual user and product rating vectors, we first build a general model of rating behavior over \emph{all} users and products, respectively.  Note that
this can be construed as an elementary user/product behavior model (we will relax the assumptions for a single model of behavior later in the section).  
For example, presume that the general pattern of rating behavior over all users follows the distribution $[0.15, 0.1, 0.05, 0.3, 0.4]$ (total proportions of 1s, 2s, 3s, 4s and 5s, respectively).  Then, we can describe this model distribution $C$ as a general trend that we expect a given user $U$'s edge-attribute (rating) value vector $f_u$ to obey, and describe the vector of $|f_u|$ rating values with respect to this model in our formulation.  
In doing so, our total encoding length in bits for each user is as follows:
\begin{align}
 L(U, M) &= L(M) \; + \; L(U|M) \\ \nonumber
\end{align}
where 
\begin{align}
\label{eqn:1}
   L(U|M) &= |f_u| \cdot \big(H(\hat{U}) + KL(\hat{U} \parallel C)\big).
\end{align}
A similar encoding cost could be written for each product. 
Following the earlier description of MDL, $L(M)$ is the cost to encode the
overall model, in this case $C$, and $L(U|M)$ is the cost to encode a
particular user's data given the model.
Here the cost for encoding the user based on the model includes the Shannon
entropy $H$ and the Kullback-Leibler divergence $KL$.  While the Shannon entropy reflects the inherent information content of the distribution $\hat{U}$, the KL divergence captures the difference between the user distribution $\hat{U}$ and the model distribution $C$:
\begin{align*}
 KL(U \parallel C) &= \sum_{i} U(i) \log_2 \frac{U(i)}{C(i)}
\end{align*}
Here both $U$ and $M$ are probability distributions over a discrete set of
outcomes, and $U(i)$ and $C(i)$ denote the probability mass associated with the outcome $i$ in each distribution.

Although this general construction described in Equation \ref{eqn:1} is required for fully encoding and reconstructing the rating vector given by a single user $U$ or given to a single product $P$ according to MDL, our goal is to be able evaluate and compare the abnormality $\delta$ of two users (and without loss of generality, products) $U_1$ and $U_2$ according to our data model, rather than evaluate the model itself.  
In this regard, the last components of the above description, $|f_{u_1}| \cdot K(\hat{U_1} \parallel C)$ and $|f_{u_2}| \cdot KL(\hat{U_2} \parallel C)$, are especially useful for abnormality comparison.  Intuitively, these terms measure the total number of \emph{extra} bits required to encode the attribute behavior of users $U_1$ and $U_2$ using a code optimized for the global model distribution $C$ respectively.  
Note that our interest in abnormality comparison no longer necessitates the use of the entropy term.  This is because the terms involving entropy are nonzero costs simply associated with the encoding of \emph{any} information irrespective of our data model.  That is, by incorporating the entropy terms, we bias our scoring function to consider inherently more expensive (random) codes as more abnormal, without any consideration of our data model -- for example, encoding a sample from the $[0.5, 0.5]$ code would necessarily be more costly than encoding a sample from the $[0.9, 0.1]$ code given their comparative entropies, regardless of the model distribution $C$.  As a result, since the terms we consider do not include entropy, we are not measuring the \emph{total} information content, but rather only the information content \emph{with respect to our model}.
Hence, we define our initial formulation $\delta_{base}$ as follows:

\begin{definition}[Base]
\small
\label{dfn:base}
Given a single edge-attribute with model distribution $C$, the base abnormality scoring function $\delta_{base}$ for node $v \in \mathcal{V}$ is defined as 
\begin{align*}
 \delta_{base}(v) = |f_v| \cdot KL(\hat{v} \parallel C)
 \end{align*}
\end{definition}
where $|f_v|$ gives the cardinality of the edge-attribute value vector $f_v$ produced from $v$'s neighboring (outgoing) edges, $\hat{v}$ gives the discrete probability distribution associated with node $v$ over the chosen attribute and $C$ gives the global discrete probability distribution of the chosen attribute over all edges.

This formulation admits two especially desirable properties:

\begin{observation}
\label{obs:1}
Given two users $U_1$ and $U_2$ where $KL(\hat{U_1} \parallel C) = KL(\hat{U_2} \parallel C)$ and $KL(\hat{U_2} \parallel C) > 0$, if $|f_{u_1}| > |f_{u_2}| > 0$, then $\delta_{base}(U_1) > \delta_{base}(U_2)$.  
\end{observation}

Observation \ref{obs:1} formalizes the intuition that given equal deviation from the model, the user who deviates on a larger scale is more surprising than the user who deviates on a smaller scale.

\begin{observation}
\label{obs:2}
Given two users $U_1$ and $U_2$ such that $KL(\hat{U_1} \parallel C) > KL(\hat{U_2}) > 0$ and  $|f_{u_1}| = |f_{u_2}|$ and $|f_{u_2}| > 0$, then $\delta_{base}(U_1) > \delta_{base}(U_2)$.  
\end{observation}

Observation \ref{obs:2} formalizes the intuition that given an equal number of ratings, the user who deviates more from the model is more surprising than the user who deviates less.

\vspace{1mm}
Note that Definition \ref{dfn:base} gives a base formulation $\delta_{base}$, for the elementary case in which we have a relation with a single, global model distribution $C$ for just a single edge-attribute.  We next relax these assumptions and discuss how to extend this formulation to more complex scenarios.  We first discuss extensions to scoring a multifaceted model in which we consider multiple model distributions for a single attribute, and next broach the topic of building a joint scoring function which can additionally incorporate multiple attributes.  Finally, we touch upon expanding these definitions to a unified scoring scheme which can handle more complex, heterogeneous graph structures with multiple relation types.  Our end goal is to devise a formulation of $\delta$ which accounts for all of these factors in ranking abnormality. 

\subsection{Handling multiple patterns of edge behavior}

It is often the case that patterns in user behavior are more granular than singular, global trends.  For example, different users may rate products in different ways.  Given that some users will be less critical and more easily satisfied than others, we may expect that some fraction of users give generally positive ratings (4s and 5s) but very few negative ratings (1s and 2s) or neutral ratings (3s).  Conversely, some users will be very difficult to please, and will heavily penalize any perceived flaws in a product by giving mostly negative and neutral ratings and very few positive ratings.  One can consider that many such latent user behaviors may exist as a result of distinct user preferences, response bias, fundamental differences in the quality of products purchased and a number of other factors.  Though our discussion here is motivated by user behavior, similar reasoning can be employed for movies, products, and other conceivable network entities across which interactions may obey multiple possible 
patterns, rather than a single standard.

In fact, $\delta_{base}$ can be extended to incorporate such a multifaceted model without much complication.  The base formulation assumes the existence of a single, global model $C$ which describes the attribute distribution over all edges.  In the user-rates-product scenario, we can consider $C$ to be a discrete probability distribution defined over the 1-5 rating values.  To capture the notion of multiple models of rating (and without loss of generality, attribute) behavior, we introduce the notation $C_{u,j}$ and $C_{p,k}$ to denote the $j$th model distribution for user ratings and the $k$th model distribution for product ratings\footnote{In general, we have for each object type $b\in B$ corresponding cluster distributions $C_{b,i}$.}, where $j \in \{1 \ldots s\}$ and $k \in \{1 \ldots t\}$ given $s$ total user rating distributions and $t$ total product rating distributions.  We can consider these as clusters which describe various modes of rating behavior.  In addition to the cluster distributions, we 
also 
define their proportions $\rho_{u,j}$ and $\rho_{p,k}$ as the fraction of user and product nodes which 
belong to the $j$th and $k$th clusters respectively -- here, we consider that a user $U$ belongs to a cluster $j$ if the Euclidean distance from $\hat{U}$ is smaller to the distribution $C_{u,j}$ than for all other clusters $\{1 \ldots s\} \setminus \{j\}$.  The analagous definition applies to a product $P$ and cluster $k$.  Note that with the introduction of such a multifaceted model, our model distribution $C$ is defined \emph{separately} for user and product ratings -- this is in constrast to the definition when we considered a single, global model.  The distinguishing factor is that in considering multiple clusters, the patterns in how users rate and how products are rated can actually differ depending on the specific edge structure of $G$.

Given the case of several underlying attribute behaviors, we face the problem of identifying abnormality as a function of multiple clusters rather than just a single one.  The abnormality of a node should also reflect to what extent its behavior fits with these various cluster distributions -- for instance, even if there are two clusters of user rating behavior, if one cluster is more widespread and characteristic of general user rating behavior than the other, this factor should be intuitively accounted for in the scoring.  To account for this concept, we introduce the following definition of the multifaceted abnormality scoring function $\delta_{mf}$:

\begin{definition}[Multifaceted]
\small
\label{dfn:multifaceted}
Given a single edge attribute and $h$ cluster distributions of type $b \in \mathcal{B}$ indicated by $C_{b,g}$ where $g \in \{1 \ldots h\}$, the multifaceted abnormality scoring function $\delta_{mf}$ for a node $v \in \mathcal{V}$ with $\Psi(v)=b$ is defined as
\begin{align*}
 \delta_{mf}(v) = |f_v| \cdot \sum\limits_{g=1}^h \big(\rho_{b,g} \cdot KL(\hat{v} \parallel C_{b,g})\big)
 \end{align*}
\end{definition}
where $|f_v|$ gives the cardinality of the edge-attribute value vector $f_v$ produced from $v$'s neighboring (outgoing) edges, $\hat{v}$ gives the discrete probability distribution associated with node $v$ over the chosen attribute, and $C_{b,g}$ and $\rho_{b,g}$ give the $g$th model distribution and proportion of the $g$th cluster respectively.

This scoring function intuitively gives the \emph{expected} number of extra bits required to encode the behavior of $v$ on a single edge attribute with respect to multiple cluster distributions.  To see this, observe that $\delta_{mf}$ is in fact the expectation over some discrete random variable $X$ with a probability mass function defined by the cluster proportions $\rho_{b,g}$ for $g \in \{1 \ldots h\}$, and outcomes defined by $\delta_{base}(v)$ for $v \in V$ and cluster distribution $C_{b,g}$.  This extension to the base formulation admits yet another desirable property:

\begin{observation}
\label{obs:3}
Given two cluster distributions $C_{u,1}$ and $C_{u,2}$ with proportions such that $\rho_{u,1} > \rho_{u,2}$ and users $U_1$ and $U_2$ such that $\hat{U_1} = C_{u,1}$ and $\hat{U_2} = C_{u,2}$, if $KL(\hat{U_1} \parallel C_{u,2}) = KL(\hat{U_2} \parallel C_{u,1})$ and $KL(\hat{U_2} \parallel C_{u,1}) > 0$ and $|f_{u,1}| = |f_{u,2}|$ and $|f_{u,2}| > 0$, then $\delta_{mf}(U_1) < \delta_{mf}(U_2)$.
\end{observation}

Observation \ref{obs:3} formalizes the intuition that in the case where two users which have no deviation from their own cluster distribution have equal deviations from the other cluster's distribution, and otherwise give an equal number of nonzero ratings, then the user who belongs to the bigger cluster is less surprising.

\vspace{1mm}
Note that by incorporating multiple patterns of edge behavior in this way, the multifaceted model inherently allow for the possibility of capturing abnormal behavior as part of the model itself.  In fact, we may find groups of users who form their own clusters based on abnormal rating patterns as a result of fraud or suspicious activity.  However, by computing the expectation over clusters using the cluster proportions as probabilities, we can still robustly identify abnormal users assuming they make up a small fraction of all users, given that they will deviate substantially from the rest of the data.  The intuition is because although they may cost few bits to encode with respect to their own abnormal cluster distribution, they will still cost many bits to store with respect to the other cluster distributions, which are weighted much more substantially due to their larger constituency in the data.

\subsection{Handling multiple edge-attributes}

We now broach the topic of building a joint abnormality function which incorporates the presence of multiple edge attributes in addition to multifaceted models on each of the individual attributes.  This is particularly useful in practical applications, where service providers collect a variety of information about each interaction.  For example, in the user-rates-product scenario, practitioners may also collect auxiliary information about the rating interaction included timestamp, rating/review text, or verification information (indicating whether the user actually purchased the product before rating it).  Each of these attributes collects information about a different aspect of the interaction which may indicate fraudulent, suspicious or otherwise anomalous but simply interesting behavior.  For example, consider a user whose given rating distribution was not itself atypical, but had a consistent inter-arrival time (IAT) of 5 seconds between ratings, meaning that each subsequent rating was given 5 seconds 
after the previous -- it is apparent in such a case that this reviewer's abnormality would not be well-indicated on the rating attribute, but would appear strongly on the temporal attribute (naturally assuming that reviews in rapid succession were not indicative of typical user behavior).  It is thus important to consider how to rank abnormality in the presence of multiple such attributes which can each be described by multifaceted models.  

There are a number of strategies we could employ for incorporating multiple attributes into the ranking context.  One strategy is to consider ranking in a subspace formulation, where we consider abnormality with respect to various subspaces of edge attributes, such as rating values, or rating values and time, or time and review text.  However, this approach introduces several problems.  Firstly, the number of subspaces grows intractably with increasing numbers of attributes -- we would have to focus on all one-attribute subspaces, two-attribute subspaces, etc.  Secondly, while considering the attributes jointly in a subspace fashion would conceivably allow for a richer data model particularly in the presence of many data points and very few attributes, it has the marked downside of sparsity issues in higher-dimensional attribute subspaces. 

A second strategy is to consider abnormality additively over each of the attributes, assuming independence.  In this approach, we compute the $\delta_{mf}$ score for each user over each attribute and simply sum the scores together.  We find that this approach offers numerous comparative advantages over the previously mentioned joint subspace method.  Firstly, instead of focusing on the combinatorial number of underlying attribute subspaces, we focus on just a single space spanned by all attributes.  This gives us a single abnormality ranking in which the top-ranking users are those who score highly in abnormality on all attributes rather than a small subset.  Furthermore, defining an additive measure of abnormality offers an attractive interpretation from the compression perspective -- since each of the $\delta_{mf}$ scores over attributes represent the expected number of extra bits to encode a user's values for that given attribute, the sum represents the expected number of extra bits to encode a user's 
values with respect to a 
joint model (though formulated independently) over all edge attributes.  The summed abnormality scores are thus naturally weighted by the deviation in terms of information content (in bits) from their respective models.  

We slightly modify our existing notation from the multifaceted (multiple clusters per attribute) model to distinguish cluster distributions between attributes $w \in \{1 \ldots y\}$ on a single relation.  Now, instead of $C_{u,j}$ and $C_{p,k}$ to denote the $j$th cluster distribution for user ratings and $k$th cluster distribution for product ratings, we write $C_{u,w,j}$ and $C_{p,w,k}$ to denote the $j$th user cluster distribution and $k$th product cluster distribution for attribute $w$, respectively.  Similarly, we write proportions as $\rho_{u,w,j}$ and $\rho_{p,w,k}$ for the proportion of the $j$th user cluster and $k$th product cluster for the $w$th attribute, respectively.  Additionally, each attribute $w$ may have a different number of user and product clusters so we write $j \in \{1 \ldots s_w\}$ and $k \in \{1 \ldots t_w\}$ where $s_w$ and $t_w$ denote the total number of user and product cluster distributions for the $w$th attribute, respectively.  Thus, we define $\delta_{ma}$ as follows:

\begin{definition}[Multi-attribute]
\small
\label{dfn:multiattr}
Given multiple edge attributes $w \in \Omega(r)$ defined on a single relation $r \in \mathcal{R}$, with $h_w$ cluster distributions of type $b \in \mathcal{B}$ respectively indicated by $C_{b,w,g}$ where $g \in \{1 \ldots h_w\}$, the multi-attribute abnormality scoring function $\delta_{ma}$ for node $v \in \mathcal{V}$ with $\Psi(v)=b$ is defined as 
\begin{align*}
 \delta_{ma}(v) = |f_v| \cdot \sum\limits_{w\in \Omega(r)} \Big(\sum\limits_{g=1}^{h_w} \big(\rho_{b,w,g} \cdot KL(\hat{v_w} \parallel C_{b,w,g})\big) \Big)
 \end{align*}
\end{definition}
where $|f_v|$ gives the cardinality of the edge-attribute value vector $f_v$ produced from $v$'s neighboring (outgoing) edges, $\hat{v_w}$ gives the discrete probability distribution associated with node $v$ over attribute $w$, and $C_{b,w,g}$ and $\rho_{b,w,g}$ give the $g$th model distribution and proportion of the $g$th cluster on the $w$th attribute respectively.

\subsection{Handling multi-relation heterogeneous graphs}

Thus far, we have built up $\delta_{ma}$ as an abnormality scoring function which handles multiple edge attributes with multifaceted models indicating various clusters of node behavior.  Now, we briefly discuss how to extend this scoring function to more complex heterogeneous schemas with multiple relation types ($|\mathcal{R}| > 1$).  Handling multiple relation types is yet another factor which can enable richer anomaly detection.  For example, consider that in our running user-rates-product scenario, we additionally incorporate a new object type of seller, and introduce a new relation user-rates-seller.  Now, one can envision a similarly motivated scenario for the multi-attribute formulation -- however, instead of a user giving typical rating values with atypical IATs, the user could now give typical rating values even with typical IATs for products but not for sellers.  Thus, considering only the user-rates-product relation for the user, we might not be able to identify a user as abnormal using the $\delta_{ma}$ 
score.  However, incorporating the secondary user-rates-seller relation, we are able to appropriately penalize the user's atypical behavior.

Fortunately, extending the formulation to handle multiple relations per object follows a very similar argument to the multi-attribute scenario where we consider handling multiple attributes per relation.  We now define a joint model on the object type which incorporates multiple relations per object, and multiple attributes per relation.  Given such a model, users who behave atypically on multiple types of interactions will be considered the most abnormal.  We can again devise an additive formulation with a minor modification to notation -- given that a user may have rated a different number of products than sellers, we use the notation $f_{u,r}$ for user $U$'s vector for relation $r$, and $|f_{u,r}|$ for the size of the attribute vector.  Similarly, we write $f_{p,r}$ and $|f_{p,r}|$ for product $P$'s vector and the associated size for relation type $r$.  Then, we define the unified heteregeneous, multi-attribute and multifaceted abnormality scoring function $\delta$ as follows: 

\begin{definition}[Unified]
\small
\label{dfn:unified}
Given multiple edge attributes $w \in \Omega(r)$ defined on multiple relations $r \in \mathcal{R}$, with $h_w$ cluster distributions of type $b \in \mathcal{B}$ respectively indicated by $C_{b,r,w,g}$ where $g \in \{1 \ldots h_w\}$, the unified abnormality scoring function $\delta$ for a node $v \in \mathcal{V}$ with $\Psi(v)=b$ is defined as 
\begin{align*}
 \delta(v) = \sum\limits_{r\in R}\Bigg(\sum\limits_{w\in \Omega(r)}\Big(|f_{v,r}| \cdot \sum\limits_{g=1}^{h_w} \big(\rho_{b,r,w,g} \cdot KL(\hat{v_w} \parallel C_{b,r,w,g})\big)\Big)\Bigg)
 \end{align*}
\end{definition}
where $|f_{v,r}|$ gives the cardinality of the edge-attribute value vector $f_{v,r}$ produced from $v$'s neighboring (outgoing) edges of type $r$. Formally, $f_{v,r}=\{ e\in E \mid v\in m(e) \wedge \Psi(e)=r  \}$. Furthermore, $\hat{v_w}$ gives the discrete probability distribution associated with $v$ over attribute $w$, and $C_{b,r,w,g}$ and $\rho_{b,r,w,g}$ give the $g$th model distribution and proportion of the $g$th cluster on the $r$th relation type respectively.

Note that the definition of $\delta$ given in Definition \ref{dfn:unified} is the final formulation of the abnormality scoring function.  From a compression perspective, it gives the expected number of extra bits required to encode a given node's edge-attribute vectors with respect to a joint model over multiple relations, multiple attributes and multiple per-attribute clusters.  In the user-product-seller scenario, we can consider that for each user, we compute deviation with respect to a joint model over the user-rates-product and user-rates-seller relations, each of which has multiple attributes (rating value, IAT, etc.) and various clusters representing patterns of behavior.  The definition is general, and extends to various node types with various numbers of relations and attributes.  

\section{Proposed Method: {\large \method}}

Thus far, we have built up both intuition and formalization for the use of $\delta$ as an abnormality score for nodes in edge-attributed graphs. 
We next describe our \method algorithm, which draws the attention of the analyst/practitioner to the nodes with the most surprising behavior in the given network.  
The pseudocode for \method is given in Algorithm \ref{alg:edgecentric} -- we describe the five associated key steps below.

\begin{algorithm}[t]
\small
\renewcommand{\algorithmicrequire}{\textbf{Input:}}
\renewcommand{\algorithmicensure}{\textbf{Output:}}
\caption{\label{alg:edgecentric} \method}
\begin{algorithmic}[1]
 \REQUIRE graph $G$
 \ENSURE sorted abnormality score vector for each node type in $G$
 \STATE For each node in $G$, aggregate attribute values from outgoing edges per-relation-type.
 \STATE Based on attribute type and range of values, discretize the space categorically for categorical attributes, and linearly or logarithmically for numerical attributes.  Bin the per-node aggregated attribute values accordingly and normalize to construct probability mass functions.
 \STATE For each node-type and attribute, cluster the vectors describing the per-attribute probability mass functions associated with each relation.
 \STATE For each node-type, compute the abnormality score $\delta$ for all nodes over associated relations and attribute clusters.
 \STATE For each node-type, sort (descending) the resulting abnormality scores and return with node indices.
\end{algorithmic}
\end{algorithm}

\newcommand{\myparagraph}[1]{{\bf #1}:}
\vspace{2mm}
\noindent \myparagraph{Step 1 -- Aggregation} For each node-type in our input graph $G$, we aggregate the attribute values over the outgoing edges from each node, for each associated relation-type.  
In our user-rates-products scenario, we have two object-types (users and products).  For this relation, we can consider two attribute types: rating values (a categorical attribute) and timestamp (a numerical attribute).  Since our relation is undirected, for each user we aggregate the attribute values for the adjacent edges, thereby collecting a vector of rating values given by the user as well as a vector of timestamps associated with the actions.  
We do the same for each object-type (in our example, both users and products).
During this aggregation step, we also collect information about attribute ranges (minimums and maximums) as well as information about each attribute type (categorical or numerical, and the more special case of temporal).  

\vspace{2mm}
\noindent \myparagraph{Step 2 -- Discretization}
Given the attribute types and ranges, 
we discretize the value space of each attribute in a principled manner.  
Categorical data is by definition discrete and thus does not need further processing.
For example, in the user-rates-product scenario, ratings take categorical values from 1-5 inclusive; each such value becomes a bin, and thus, each user becomes a histogram of $d = 5$ entries.
For numerical attributes, the discretization process requires more sophistication.  We propose an adaptive binning approach as follows: if the maximum value of the attribute is an order of magnitude (at least 10 times) larger than the minimum value, we space the bin markers logarithmically into some prespecified number of $d$ bins ($d = 20$ in our experiments).  Otherwise, we space bin markers linearly in the same fashion.  Logarithmic binning addresses issues associated with sparsity and scale insensitivity which arise when trying to bin a large range of values linearly.  Conversely, when the magnitude of values is not large, linear binning tends to be sufficient in approximating the data distribution.  

\vspace{2mm}
\noindent \myparagraph{Step 2' -- Discretization of temporal data}
One notable exception to this treatment of numerical attributes is the handling of temporal data, e.g. timestamps of ratings.  Binning the raw timestamps is not useful given the monotonically increasing and uninformative nature of the absolute values.  However, the interarrival time (IAT), or time between subsequent timestamps, is a more interesting attribute for consideration.  Firstly, it can be computed very easily as a first-difference of the vector of interaction timestamps for a given node.  Secondly, trends in IATs can reflect certain informative aspects of honest user behavior (patterns in frequency of ratings and general service usage) as well as fraudulent or abnormal user behavior (fixed periodicity ratings, with short IATs).  For example, a user giving ratings every 5 seconds may be indicative of bot behavior.  Similarly, a product receiving many ratings every 5 seconds may be indicative of a seller trying to game the rating system by purchasing fake ratings from a botnet operator.  Using IAT as an attribute instead of raw timestamps is attractive for these reasons.  Upon computing the interarrival time vector for temporal numerical attributes, the binning process follows as described in Step 2.

\vspace{2mm}
\myparagraph{Step 3 -- Clustering}
After binning the per-node attribute values and normalizing to construct the appropriate probability mass functions, we cluster the vectors describing the probability masses as a number of $d$-dimensional points (where $d$ representative the number of discrete histogram bins).  
Though any clustering algorithm could be used for this purpose, 
we use and recommend
$X$-means~\cite{pelleg2000xmeans}, because it
automatically chooses the number of clusters in a principled manner by optimizing Bayesian Information Criterion (BIC).
The centers of the resulting clusters are $d$-dimensional
probability mass functions themselves, which we use as the 
cluster distributions.  We can then compute cluster proportions by empirically assigning the input points to clusters in traditional fashion (smallest $L^2$ distance).

\vspace{2mm}
\noindent \myparagraph{Step 4 -- Scoring}
Given the cluster distributions across all attributes and node-types over the respective relations, we now compute the abnormality score $\delta(v)$ for each node $v \in \mathcal{V}$ according to Definition \ref{dfn:unified}.  
For each object-type, and over each of the attributes on associated relations, we {\it additively} compute the abnormality score in terms of the expected cost in extra bits with respect to the attribute cluster distributions.  

\vspace{2mm}
\noindent \myparagraph{Step 5 -- Ranking}
Finally, we sort the scores for each object-type in a descending fashion and return the ranking with associated node indices to the practitioner.  This effectively routes practitioner attention to the most abnormal nodes for each of the object-types in the graph (users, products, etc.), with respect to encoding cost over a joint model composed of independent edge-attribute models.  This information can then be leveraged for further investigation.

\section{Experimental Analysis}

In this section, we evaluate \method and aim to answer the following questions: what kinds of edge-attribute behavior do we observe in real-world graphs?  Is \method practically effective in finding abnormally-behaving nodes by leveraging this information?  Finally, is \method scalable?

\subsection{Datasets and Experimental Setup}

\begin{table}[t!]
\centering
\scriptsize
\caption{Datasets used for empirical analysis}
\label{tbl:datasets}
    \begin{tabular}{lll}
    \toprule
    {\bf Graph}               & {\bf Nodes} & {\bf Edges}                                                                                                                         \\ \midrule
    \flipkart \cite{flipkart} & 1.1M users, 545K products & 3.3M ratings \\
    \swm \cite{akoglu2013opinion} & 964K users, 15K applications & 1.1M ratings  \\ 
    \amazon \cite{mcauley2015inferring} & 1.8M users, 252K products & 3.0M ratings \\ \bottomrule
    \end{tabular}
\end{table}

For our evaluation, we apply \method to 3 real-world graphs with various edge-attributes.  The datasets are summarized in Table \ref{tbl:datasets} and described in further detail below.

\vspace{1mm}
\noindent \textbf{Flipkart:} The \flipkart dataset contains information about reviews and ratings in the Flipkart e-commerce network which provides a platform for sellers to market products to customers.  It contains roughly 3.3 million ratings given by 1.1 million users to 545 thousand products over the timespan of August 2011 to January 2015. 

\vspace{1mm}
\noindent \textbf{Software Marketplace:} The \swm dataset contains information about purchases in an anonymous online marketplace which allows customers to purchase software applications.  The data for this marketplace was originally collected and used in \cite{akoglu2013opinion}.  It contains over 1.1 million ratings given by 964 thousand users to 15 thousand applications over the timespan of April 2008 to June 2012.

\vspace{1mm}
\noindent \textbf{Amazon Health and Personal Care:} The \amazon dataset is publicly available.  It contains information about online purchases of health and personal care products in the Amazon e-commerce network, which also provides a platform for sellers to market products to customers.  It contains roughly 3.0 million ratings given by 1.8 million users to 252 thousand products over the timespan of May 1996 to July 2014.  

\vspace{1mm}
Our code for \method is primarily implemented in Python, with use of NumPy library.  Additionally, we use the C++ $X$-means implementation given in \cite{pelleg2000xmeans} for the clustering task (step 3 in Algorithm \ref{alg:edgecentric}).  The $X$-means approach constructs a $kd$-tree for efficient neighbor querying and automatically infers the appropriate number of data clusters and their locations by optimizing the BIC measure.  We use the suggested parameters.  For our experiments, we use a machine with 32 Intel Xeon 8837 CPU cores running at 2.67GHz each and 1TB of total RAM.  

\subsection{Findings on Flipkart}

\begin{figure}[t!]
\centering
\includegraphics[width=0.49\textwidth]{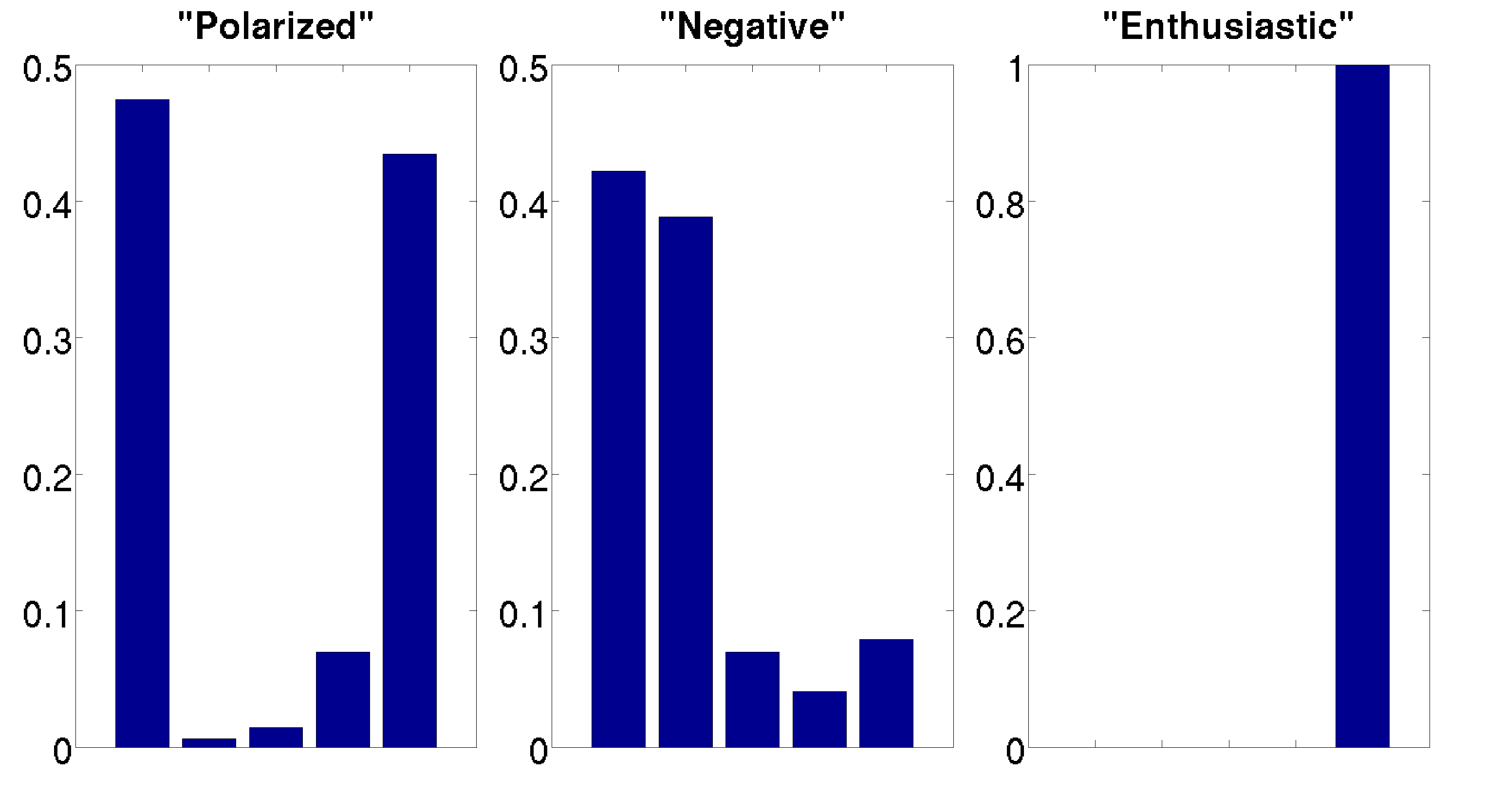}
\caption{\label{fig:userratingclusters} {\bf Discovered popular user-rating patterns.}  Here, we show several cluster distributions and associated probability masses for user ratings on the \flipkart dataset -- bins correspond to 1-5 stars.}
\end{figure}

\begin{figure}[t!]
\centering
\includegraphics[width=0.49\textwidth]{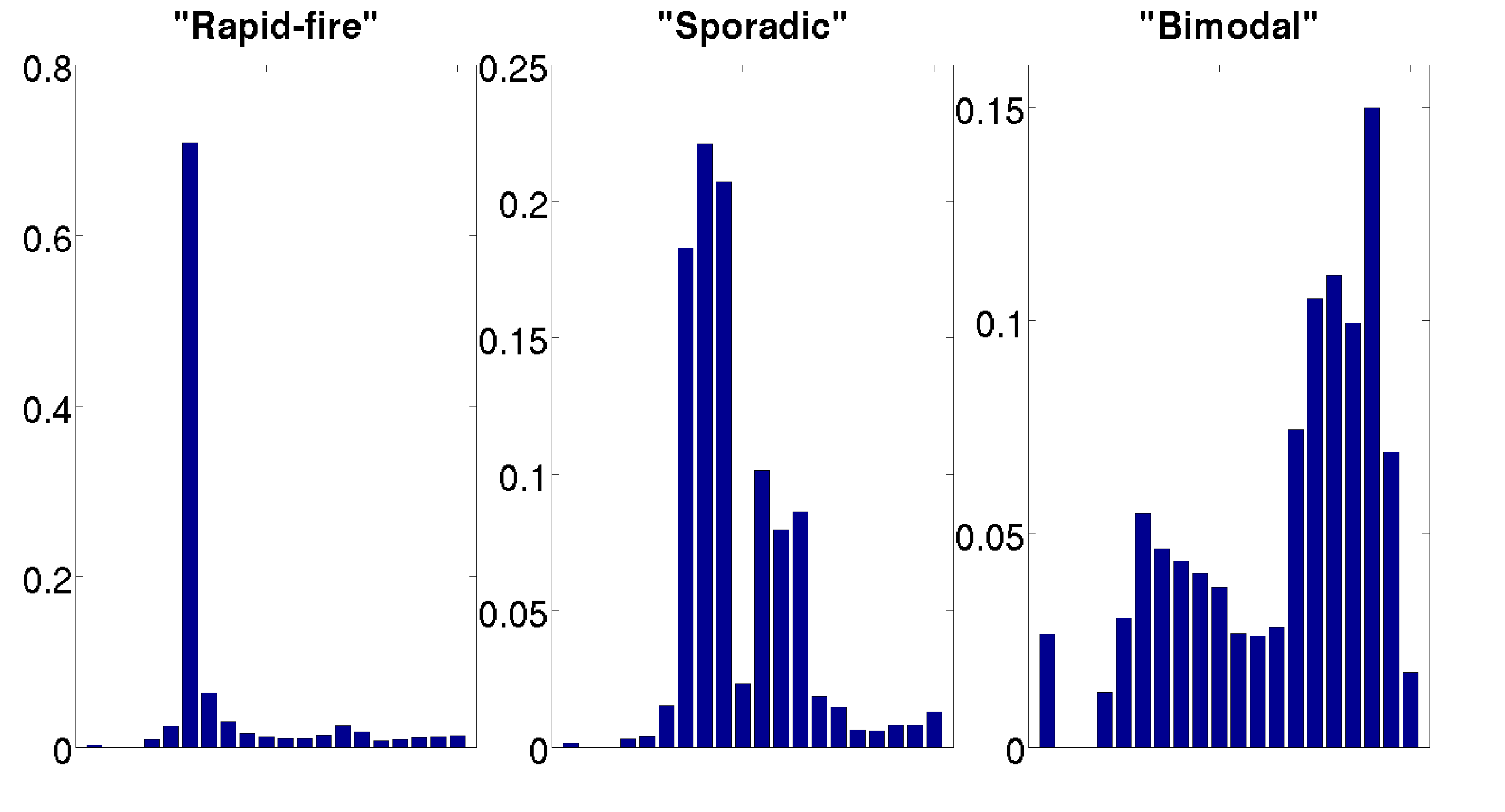}
\caption{\label{fig:usertimeclusters} {\bf Discovered popular rating frequency (IAT) patterns.}  Here, we show several cluster distributions and associated probability masses for interarrival times between ratings on the \flipkart dataset -- bins correspond to logarithmically discretized interarrival times ranging from orders of seconds to years.}
\end{figure}

In our analysis on the \flipkart dataset, we constructed a single relation (user-rates-product) on which we had one categorical attribute (rating in stars from 1-5) and one temporal numerical attribute (UNIX timestamp of rating).  Thus, we ranked abnormality of users with respect to their rating and IAT behavior.  

Figures \ref{fig:userratingclusters} and \ref{fig:usertimeclusters} show the probability mass functions corresponding to the distributions that we found as a result of clustering the user edge-attribute data.  Figure \ref{fig:userratingclusters} shows several interesting rating patterns we discovered from the 17 total clusters produced from the $X$-means process: \emph{polarized}, \emph{negative} and \emph{enthusiastic} users.  \emph{Polarized} users give mostly 1 star and 5 star ratings, with very few middle-ground ratings -- this can correspond to the natural tendency to either love or hate a product, or result from fraudulent users who aim to popularize all the products of a single seller, and defame the competitors.  \emph{Negative} users give mostly 1 or 2 star ratings -- we conjecture this is mostly a consequence of response bias, where users are sharing their opinions only because they are especially displeased with a product.  Finally, \emph{enthusiastic} users give only 5 star ratings and none others -- this is suggestive of strong response bias or blatantly fraudulent behavior (especially when the user gives many such ratings).  We additionally find isolated clusters for users who give only ratings of a single star outcome (1-5) -- these \emph{single-minded} are particularly prevalent in the data, given the large number of low-activity users who rated only one or a few products since inception.  The presence of these behaviors in various proportions of the data then informs the computation of the abnormality scores and \method rankings for individual users.  

Figure \ref{fig:usertimeclusters} shows several IAT patterns (indicating rating frequency), selected from 17 total clusters produced from the $X$-means process: \emph{rapid-fire}, \emph{sporadic} and \emph{bimodal} users.  The bins are discretized logarithmically, ranging from orders of seconds to years (normally the case for users who rate only a few products in total, with a large gap between subsequent uses of the Flipkart platform).  \emph{Rapid-fire} users are the most blatantly suspicious -- these users almost exclusively give ratings with seconds between subsequent ones.  This type of behavior is almost guaranteed to be fraudulent and does not correspond with any intuition of real human behavior.  Conversely, \emph{sporadic} users' behavior is far more in-line with human intuition.  These users mostly give ratings several weeks to months apart.  Very few ratings are given with shorter IATs, indicating that the users mostly rate single items upon purchase, and purchase only sporadically (mobile phones, birthday presents, holiday gifts, etc.)  Lastly, \emph{bimodal} users behave bimodally, in that they occasionally spend weeks to months without rating a product, but often have periods of frequent activity on the order of multiple ratings (purchases) in days to weeks.  Notice that the probability mass for the users in this cluster is distributed across almost all orders of IAT, with most of the mass concentrated in the days to weeks range, suggesting that the users are engaged with the Flipkart service and give ratings frequently (presumably because they also purchase products frequently).  However, a non-trivial amount of the mass is distributed between shorter timeframes, indicating that the users rate multiple products in a single sitting (likely due to the purchase and resulting receipt of several products at the same time).  Similar bimodal behavior in IAT has also been observed in \cite{ferraz2015rsc} with respect to comments on Reddit and tweets on Twitter.

\begin{figure}[t!]
\centering
\includegraphics[width=0.5\textwidth]{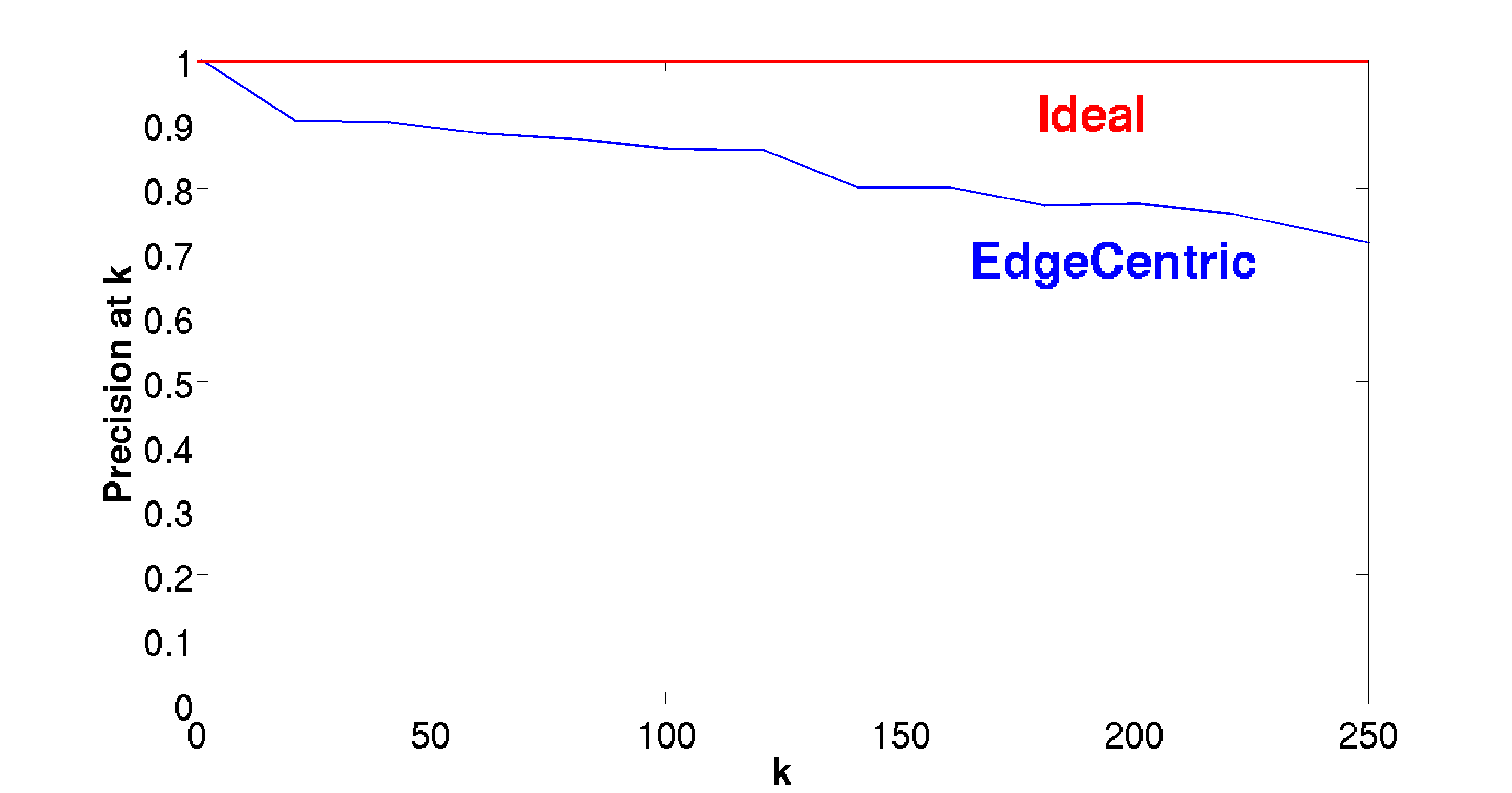}
\caption{\label{fig:precisionatk} {\bf \method finds fraudulent accounts on Flipkart with high precision.}  Here, we show the precision@$k$ for various values of $k$ ranging from 1 to 250, based on hand-labeled data from domain experts at Flipkart.}
\end{figure}

Upon applying \method to this dataset, we provided a list of the 250 most abnormal accounts to domain-experts at Flipkart who investigated and labeled these users individually according to various criteria involving the user's review-text, rating distributions and frequencies.  Figure \ref{fig:precisionatk} shows the precision at $k$ (P@k) for a spread of $k$ values over this range of 250 users, indicating positive results of almost 0.9 precision over the top 50 users, and over 0.7 precision over the top 250 users -- recall results are not available given the unbounded number of false negatives and lack of ground-truth labels for all users.  These are substantial findings for Flipkart.  One common pattern that the domain-experts found was that most of the users who were labeled as fraudulent were individual accounts who are either spamming 4/5 star ratings to multiple products from a single seller (boosting the seller's ratings), or spamming 1/2 star ratings to multiple products from another seller (defaming the competition).  We further found that the most abnormal user had given 3692 5 star ratings with an average IAT of just a few seconds.  

\subsection{Findings on Software Marketplace}

On the \swm dataset, we constructed a single relation (user-rates-application) on which we had one categorical attribute (rating in stars from 1-5).  Thus, we ranked abnormality of users with respect to their rating behavior.  We do not show the clustered rating behavior in interest of space limitations, but note that similar behaviors can be observed in this dataset in terms of polarized raters, ``single-minded'' raters, almost exclusively positive and almost exclusively negative raters, etc. as in Figure \ref{fig:userratingclusters}.

We find that the users with the highest scores according to our \method approach have very spammy behavior.  The most abnormal user in this dataset had given a single 1 star rating to one application, along with 186 5 star ratings to a different application.  The accompanying reviews had very high textual similarity and included quotes like

\vspace{3mm}
\begin{compactitem}
\small
\item {\tt ``Awesome!!!,Get this app now and earn points for a \$10 gift card.''}
\item {\tt ``Try the app today and you will be amazed of how much money you can make with it......''}
\item {\tt ``Awesome App!!!! FREE money ,The app is great to earn points for FREE money. Get it today!''}
\end{compactitem}

\vspace{3mm}
Another of the highly-ranked users had given 107 5 star ratings to a single application, spamming the following review:

\vspace{3mm}
\begin{compactitem}
\small
\item {\tt ``Great app,Just great! Enter code:    [redacted] To win even more points!!!!!!!!!!!!!!!!!!!!''}
\end{compactitem}
\vspace{3mm}

In fact, most of the top 20 users which we manually checked accompanying reviews for posted repetitive, spammy text in addition to highly skewed rating patterns.  Usually, the review text promoted the application, included personalized codes which the reviewers claimed would give the potential customer free points, gift cards or money, or were generally characteristic of meme-like and information-free content.  We additionally found correspondences between the codes reviewers asked customers to use and the reviewer's own usernames, suggesting that the code actually gave the reviewer some associated perk rather than the customer.  This is in line with the common marketing practice to incentivize existing customers to attract more potential customers.   Unfortunately, we are unable to check for ground-truth with the service providers.

\subsection{Findings on Amazon Health and Personal Care}

For the \amazon dataset, we constructed a single relation (user-rates-product) on which we have one categorical attribute (rating in stars from 1-5) and one temporal numerical attribute (UNIX timestamp of rating).  Thus, we ranked abnormality of users with respect to their rating and IAT behavior.  Clustered rating behavior is not shown in interest of space limitations, but we find the same overarching patterns as in Figures \ref{fig:userratingclusters} and \ref{fig:userratingclusters}. 

Users with abnormal rating behavior need not be examples of fraudsters or spammers, as in the \flipkart and \swm cases.  In the \amazon dataset, we observe that the top 5 users with the highest scores are in fact reviewers with high-status badges.  These users do not only review health and personal care products, but rather review a very large number of products across many categories.  These products are often provided to the top reviewers free of charge in return for a rating and review.  Good evaluations from these reviewers are seen as a status symbol for the product.  

The most abnormal user according to \method, a reviewer by the name \emph{C. Hill}, had rated 348 health and personal care products with 317 of the ratings being 4/5 stars.  A quick look at his Amazon review profile indicates that he is a Hall of Fame reviwer who has rated 5641 total products and is rank 172 on Amazon's top reviewer ranking list.  Each of the given reviews are similarly structured and content-rich.  Many of them include the following statement: ``Note: sample unit provided for reviewing purposes.''  Another highly ranked user by \method, \emph{Margaret Picky}, has ended almost all of her recent reviews with ``[Seller] provided [Product] for evaluation and review.''  At the time of data collection, the user had rated 285 health and personal care products with 262 4/5 star reviews, of which almost 80\% were 5 stars.  Interestingly, many reviewers who receive products for evaluation tend to give abnormally many more positive ratings than other users.

Several of the other top reviewers have similar badges, indicators, and review styles -- for example, 2 of the top 5 reviewers have \emph{Vine Voice} badges, indicating that the customer is a member of the invitation-only \emph{Amazon Vine Voice} program which gives reviewers advance access to not-yet-released products for the purpose of writing reviews.  \cite{npramazon} provides further details from one such top-reviewer who details his experiences in having received thousands of dollars of free products from Amazon sellers over the years for evaluation purposes.

It is comparatively difficult to make claims about the abnormality of products, from a legitimacy standpoint.  This is both because different products have inherent differences in quality and frequency of purchase, and also spammers rating the product very highly or poorly do not necessarily imply the legitimacy or illegitimacy of a product or its seller.  However, \method does in fact find abnormal products in the \amazon dataset.  Interestingly, the two most highly-ranked abnormal products were two digital bathroom weighing scales from the \emph{EatSmart} product line -- the higher-ranked of these is the \#1 best-seller in the Amazon \emph{Digital Bathroom Sales} category, with 19,593 reviews, of which 80\% are 5 star and 10\% 4 star ratings.  This is an extremely strongly rated product in comparison with others in the dataset, and also rated with very high frequency (short IAT) given the product's popularity.  Several of the other products which were highly ranked as abnormal by our approach are very 
highly-recommended fitness products including the \emph{BlenderBottle} and \emph{FitBit}, which have 91\% and 85\% 4/5 star ratings and 10,820 and 10,260 customer reviews respectively.

\subsection{Scalability}

\begin{figure}[th!]
\centering
\includegraphics[width=0.5\textwidth]{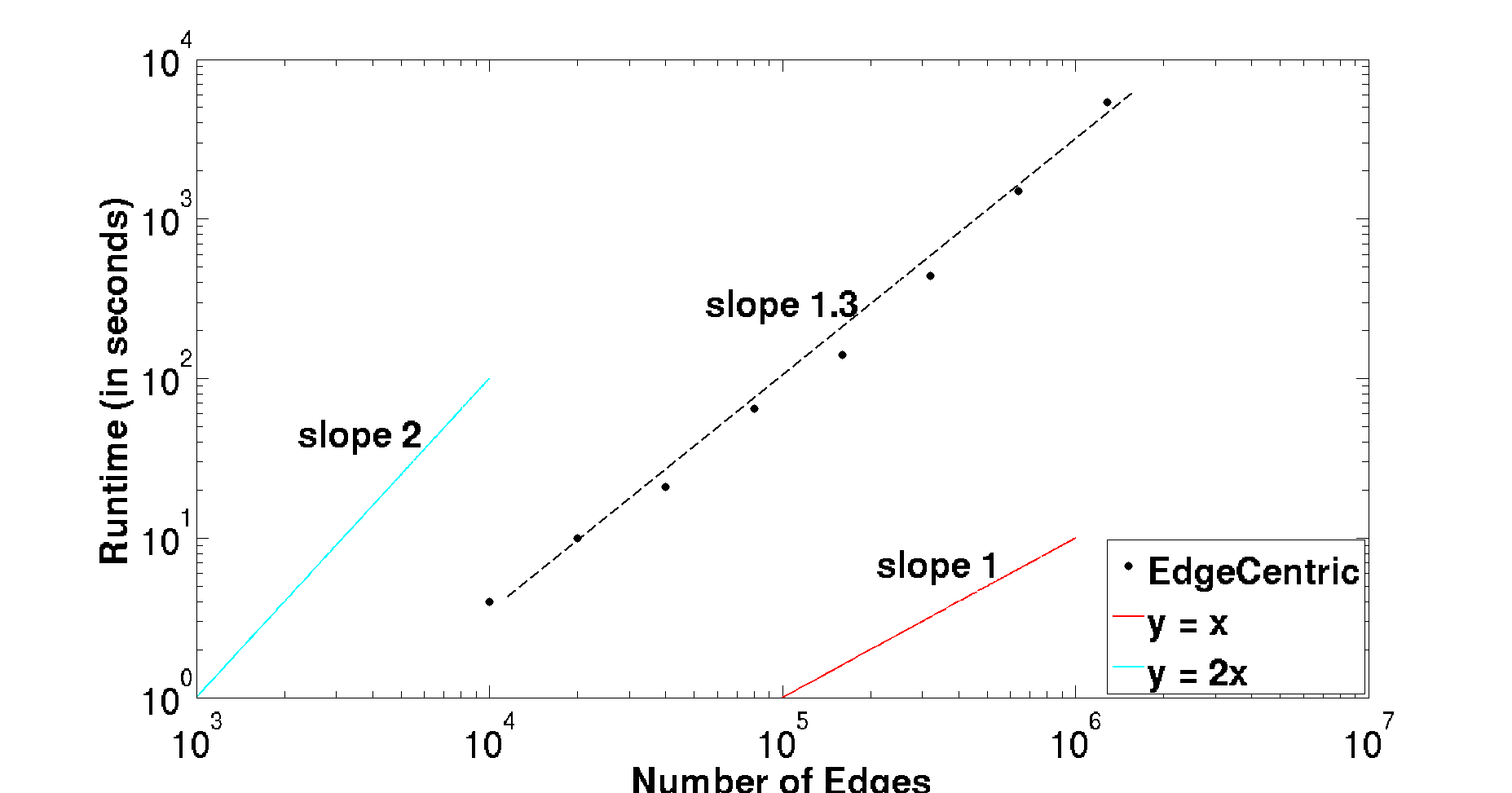}
\caption{\label{fig:scalability} {\bf \method is scalable.}  Here, we show \method's runtime on induced subgraphs of the \amazon dataset, up to 1.3M edges in size.}
\end{figure}

Finally, we show that \method is scalable on real-world graphs.  Figure \ref{fig:scalability} shows the runtime in seconds on various induced subgraphs from the \amazon dataset.  The time-complexity of \method is roughly $O(|\mathcal{E}|d \, + \, |\mathcal{V}|kdi)$ for a single attribute, where $|\mathcal{E}|$ is the number of edges (nonzeros) in $G$, $d$ is the dimensionality of the attribute, $|\mathcal{V}|$ is the number of nodes in $G$, $k$ is the number of data clusters and $i$ is the number of iterations used by the clustering algorithm.  The first term reflects the cost of the binning process for the attributes, into $d$ bins, or dimensions.  The latter term approximately reflects the cost of clustering the data using the $X$-means algorithm, which is shown to scale comparably better than Lloyd's iterative refinement $k$-means algorithm. 

\section{Conclusion}

In this work, we broach the issue of detecting anomalies in large, edge-attributed real-world graphs, which are commonplace in modern e-commerce platforms, social networks and other web services.  Specifically, we first formalize the problem of detecting anomalous nodes in graphs as a ranking problem, in which we aim to score nodes based on the abnormality of their edge behavior.  To this end, we first build up the intuition of using information theoretic principles to quantify abnormality with respect to deviation from typical behavior in a data-driven fashion, and show extend this formulation in the presence of multiple user behaviors, multiple edge-attributes and complex heterogeneous graphs.  We then introduce the \method approach to leverage this formulation to practically detect anomalies in real-world graphs.  Finally, we show substantiating results including high precision (0.87 over the top 100 users) on the Flipkart e-commerce platform, practical scalability and interesting observations on atypical user behavior gleaned from successfully applying our method to several large, existing real-world networks.
%

\section{Acknowledgements}
This material is based upon work supported by the National Science Foundation under Grant Nos. IIS-1217559, CNS-1314632 and DGE-1252522. Prepared by LLNL under Contract DE-AC52-07NA27344.   Any opinions, findings, and conclusions or recommendations expressed in this material are those of the author(s) and do not necessarily reflect the views of the National Science Foundation, DARPA, or other funding parties. The U.S. Government is authorized to reproduce and distribute reprints for Government purposes notwithstanding any copyright notation here on.

%

\bibliographystyle{abbrv}
\bibliography{bib/neil}  

\end{document}